\documentclass[seceq]{ptptex}
\usepackage{amsmath}
\usepackage{amssymb}
\usepackage{cite}
\usepackage{epsfig}
\usepackage{cancel}
\usepackage{flafter}

\newcommand{\be}{\begin{equation}}
\newcommand{\ee}{\end{equation}}
\def\spa#1.#2{\left\langle#1\,#2\right\rangle}
\def\spb#1.#2{\left[#1\,#2\right]}
\def\spaa#1.#2.#3{\langle\mskip-1mu{#1}
                  | #2 | {#3}\mskip-1mu\rangle}
\def\spbb#1.#2.#3{[\mskip-1mu{#1}
                  | #2 | {#3}\mskip-1mu]}
\def\spab#1.#2.#3{\langle\mskip-1mu{#1}
                  | #2 | {#3}\mskip-1mu\rangle}
\def\spba#1.#2.#3{\langle\mskip-1mu{#1}^+
                  | #2 | {#3}^+\mskip-1mu\rangle}
\def\spav#1.#2.#3{\|\mskip-1mu{#1}
                  | #2 | {#3}\mskip-1mu\|^2}
\def\jc#1.#2.#3{j^{#1}_{#2#3}}
\def\blfootnote{\xdef\@thefnmark{}\@footnotetext} 

\def\bea{\begin{eqnarray}}
\def\eea{\end{eqnarray}}

\title{Forward physics hard processes and saturation: theory and phenomenology review.
}

\author{
Krzysztof \textsc{Kutak}%
}

\inst{
Instytut Fizyki Jadrowej im. Henryka Niewodnicznskiego Polskiej Akademii Nauk.\\
Radzikowskiego 152, Krakow, Poland.
}

\abst{
Selected aspects of the theory and phenomenology of production of forward jets at the LHC and RHIC are overviewed. In the theory part we discuss basics of frameworks originating from the BFKL dynamics while in the phenomenology part we present results for forward-central jet correlations at the LHC and forward di-jet production at RHIC.}

\begin{document}

\maketitle

\section{Introduction}
Physics in the forward region at hadron colliders 
is traditionally dominated by soft particle production. However, with the LHC and recent results from RHIC on d-Au scattering \cite{Braidot:2010ig,Braidot:2011zj,Adare:2011sc,Meredith:2009fp}
forward physics  phenomenology turns  into  a  largely new 
field~\cite{Grothe:2009nk,Jung:2009eq,d'Enterria:2009fa}   involving 
both soft and hard production processes, 
because of  the phase space opening up at  high center-of-mass energies.  
At the LHC owing to the unprecedented reach in rapidity of  the experimental instrumentation, it becomes possible to carry out a program of high-p$_{\rm{T}}$ physics in the forward region.  
 Apart from the purely QCD relevance forward jet  production   enters 
the LHC physics program   in  an  essential way for new particle searches, e.g. 
in  vector boson fusion search channels or for the Higgs boson~\cite{VazquezAcosta:2009sx}. 
Another area  of  potential interest  in forward physics employs  
  near-beam proton taggers:  this  will enable  
studies to be made in the 
central     high-p$_{\rm{T}}$ 
production mode with forward protons, which can  be used 
 for  both standard-candle~ and 
discovery physics \cite{Heinemeyer:2010fk}. 
In addition to collider physics applications, measurements of 
forward particle production 
at the LHC  will serve as  input   to   the modeling of high-energy air showers in cosmic 
ray experiments~\cite{arXiv:1101.5596}. 
The forward  production of high-p$_{\rm{T}}$ particles brings jet physics into a 
region  characterized  by  multiple energy scales and asymmetric parton kinematics. 
In order to account for multi jet rates in this multi-scale region, it is necessary to use formalisms which go beyond fixed order and include perturbative QCD resummations. The  approaches which we are going to overview here are High Energy Factorization (HEF)~\cite{Catani:1990eg} and the HEJ \cite{arXiv:1101.5394,arXiv:1104.1316,arXiv:0910.5113} framework, for other approaches we refer the reader to \cite{hep-ph/0603175,arXiv:1102.1672}. 
Another aspect of forward jet physics is that it probes at least one of the parton density at values of $x$ low enough to study saturation effects, a phenomenon for which there is growing evidence \cite{arXiv:1005.4065,Dumitru:2010iy,Praszalowicz:2011tc}, more extensively.
\section{Hard processes in p-p scattering}
In this section we review selected frameworks for  calculations of observables characterizing the production of jets in proton-proton
 collisions with the condition that the rapidity gap between the produced jets is large. In particular we focus on the production of a forward jet associated with a central jet.  
In a proton collision process in which a forward jet is associated with a central jet,
one collimated group of high p$_\perp$ hadrons continues along the direction of one
of colliding protons - forward detector region, while another group heads toward
central region.
The high p$_\perp$ production at  microscopic level can be understood as originating from collision of
two partons where one  of them which is almost on-shell carries a large longitudinal momentum fraction $x_2$ of
mother proton ($p_2$) while the other one carries a small longitudinal momentum fraction $x_1$
of the other proton ($p_1$) and is off-shell.\\
\subsection{High Energy Factorization (HEF)}
One of the frameworks to describe forward jets is provided by
High Energy Factorization which was derived after the observation of gluon exchange dominance at high energies.
Similarly to collinear factorization it decomposes the cross-section
into parton density functions characterizing incoming hadrons
at  fixed transverse momentum, and perturbatively calculable matrix elements.
However, apart from large logarithms of hard scale it also resums large logarithms coming from energy ordering. The formula for high energy factorization takes the form:  
\bea
  \frac{d\sigma}{dy_1dy_2d^2p_{1t}d^2p_{2t}} 
  & = &
  \sum_{a,b,c,d}\int\frac{d^2k_1}{\pi}
  \frac{1}{16\pi^2 (x_1x_2 S)^2}
  \overline{|{\cal M}_{ab\to cd}|}^2\\ 
  & &
  \hskip 30pt
  \times\ \delta^2(\vec k_1 - \vec p_{3t}-\vec p_{4t}){\cal A}_{a/A}^*(x_1,k_1^2,\mu^2)\,  f_{b/B}(x_2,\mu^2)\,\nonumber\\
 & &
  \hskip 30pt \times\frac{1}{1+\delta_{cd}}\nonumber,
\eea
where $k_1\equiv|{\bf k_1}|$ and $p_{3t}$ and $p_{4t}$ are
transversal momenta of final state partons.
The function ${\cal A}_{a/A}^*(x_1,k_1^2,\mu^2)$ is the unintegrated gluon distribution which is a solution to high 
energy factorisable evolution equations like BFKL, CCFM or BK, and $f_{b/B}(x_2,k_2^2,\mu^2)$ is the integrated 
parton density which is a solution of the DGLAP equation.
They describe distributions of transversal and longitudinal momenta of partons in the incoming protons
 $A$ and $B$ respectively. The sum is made over all flavors of initial and final state partons. 
The matrix elements relevant for high energy factorization describe a hard
subprocess where at least one of the incoming partons is off mass shell. They are
calculated by applying the high-energy eikonal
projectors to scattering
amplitudes ${\cal M}$.
In reference \cite{arXiv:0908.0538}  matrix elements relevant for forward
jets phenomenology have been calculated, in fully exclusive form. The framework of high energy factorization has been implemented in the multi-process Monte Carlo event generator CASCADE \cite{arXiv:1008.0152}. 
\label{sec:trv}
\begin{figure}[t!]
  \begin{picture}(30,30)
    \put(-90, -210){
      \includegraphics{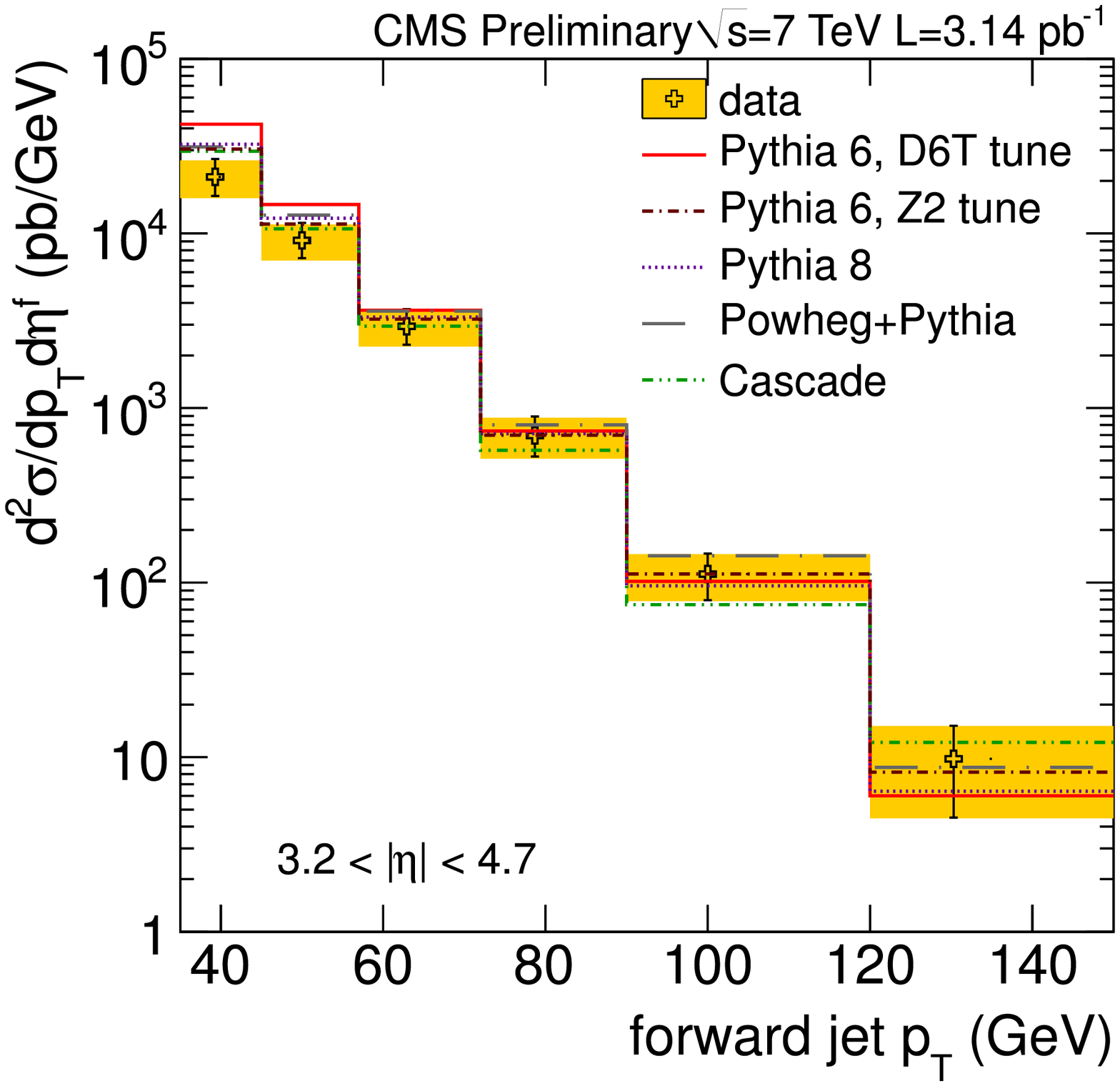}
    }
    \put(130, -210){
      \includegraphics{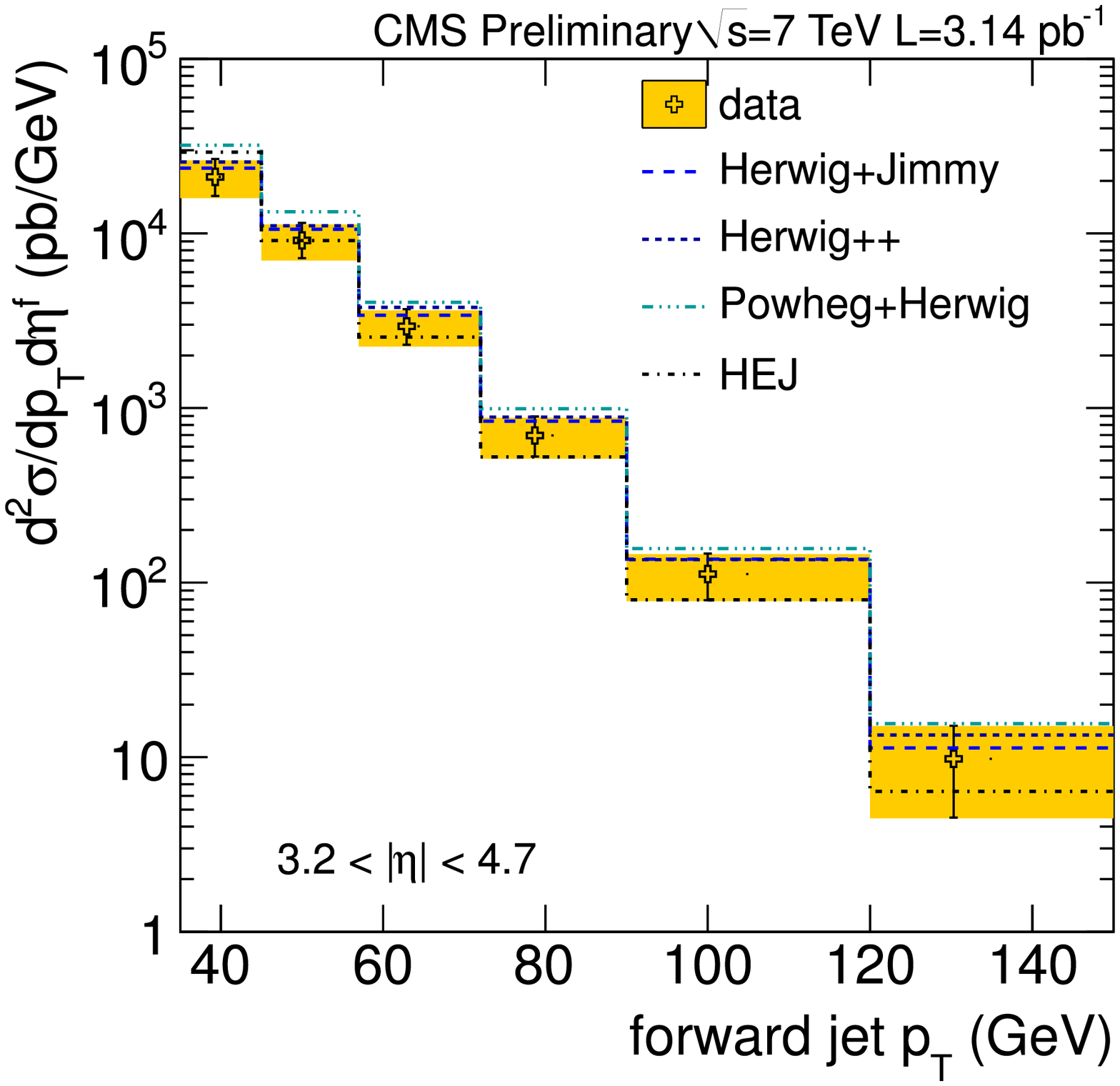}
    }

     \end{picture}
\vspace{7cm}
\caption{\em \small Transversal momentum spectra of produced jets at total collision energy $\sqrt s=7\,TeV$ with requirement 
that p$_\perp\!>\!20\,GeV$. Spectrum of forward jets (left);  
spectrum of central jets (right), the data points are the CMS measurement\cite{CMS}.}
\label{fig:transversal}
\end{figure}
\subsection{High energy jets (HEJ)}
Another framework with its origin in high energy resumation is High Energy Jets (HEJ) \cite{arXiv:1101.5394,arXiv:1104.1316,arXiv:0910.5113}.
It addresses the description of multiple hard perturbative corrections in both the (low) fixed-order and in the parton
shower formulation. The perturbative description obtained with HEJ
reproduces, to all orders in QCD, the high energy limit for both real and virtual
corrections to the hard perturbative matrix element. 
The basic idea of this approach relies on similar observation as in HEF that the dominant contribution at highest energies is due 
to the exchange of a gluon in the $t$ channel.
In this approach the limit of pure $N$-jet amplitudes for large invariant mass between each
jet of similar transverse momentum is described by the
FKL-amplitudes\cite{Fadin:1975cb,Kuraev:1976ge}, which are the foundation
of the BFKL framework\cite{Balitsky:1978ic}. The physical picture arising
from the FKL amplitudes is one of effective vertices connected by $t$-channel
propagators. In this respect this framework needs less approximation than the BFKL equation.
The virtual
corrections are approximated with the Regge gluon trajectory for the
$t$-channel gluon propagators. The final result is a formalism which provides a good approximation
order-by-order to the full QCD results, and is fast enough to
evaluate all-order results for the amplitudes which can be explicitly
constructed and integrated over the $n$-body phase spaces.
The cross section in the HEJ framework takes the form:
\begin{align}
  \begin{split}
    \label{eq:dijetreg}
    \sigma_{qQ\to2j}=&\sum_{n=2}^\infty\
    \prod_{i=1}^n\left(\int_{p_{i\perp}=0}^{p_{i\perp}=\infty}
      \frac{\mathrm{d}^2\mathbf{p}_{i\perp}}{(2\pi)^3}\ 
      \int \frac{\mathrm{d} y_i}{2}
    \right)\
    \frac{\overline{|\mathcal{M}_{\mathrm{HEJ}}^\mathrm{reg}(\{ p_i\})|}^2}{\hat s^2} \\
    &\times\ \ x_a f_{A,q}(x_a, Q_a)\ x_2 f_{B,Q}(x_b, Q_b)\ (2\pi)^4\ \delta^2\!\!\left(\sum_{k=1}^n
      \mathbf{p}_{k\perp}\right )\ \mathcal{O}_{2j}(\{p_i\}).
  \end{split}
\end{align}
where $|\mathcal{M}_{\mathrm{HEJ}}^\mathrm{reg}(\{ p_i\})|$ is the regularized matrix element in which real contributions are built from effective vertices and the virtual contributions originate from the gluon trajectory. The collinear parton densities are $x_a f_{A,q}(x_a, Q_a), x_2 f_{B,Q}(x_b, Q_b)$, and the function $\mathcal{O}_{2j}(\{p_i\})$ is equal to one if all final state particles represent jets, otherwise it gives zero.

An interesting observable of the forward central jet production process to which both of the presented frameworks have been applied and have been compared to data, is the $p_T$ 
spectrum of produced forward and central jets see Fig.~\ref{fig:transversal}. This observable is useful for example in testing and pushing further the development of hard scale dependent unintegrated gluon densities. This is necessary first of all because so far the unintegrated gluon densities were applied mainly to processes at low scales and secondly because the hard scale is not introduced in a unique manner.
\section{Saturation and hard processes in hadron-nuclei scattering}
\subsection{Saturation at RHIC}
Perturbative Quantum Chromodynamics (pQCD) at high energies generates a scale called the saturation scale $Q_s$.
The need for the existence of such a scale originates back to investigations of unitarity violation \cite{Gribov:1984tu} by linear 
equations 
of pQCD \cite{Kuraev:1977fs,Balitsky:1978ic}.
To resolve this problem, higher order perturbative corrections of nonlinear type within the Balitsky-Kovchegov, CGC/JIMWLK  
framework (i.e. nonlinear modifications to summation of logarithms of the type $\alpha_s^n(\ln s)^n$) were considered 
\cite{Balitsky:1995ub,Kovchegov:1999yj,Kovchegov:1999ua,JalilianMarian:1997gr,JalilianMarian:1998cb,Kovner:2000pt,Weigert:2000gi,
Iancu:2000hn,Ferreiro:2001qy}. These unitarity corrections have a clear physical meaning for they require gluons to recombine. The saturation scale depends on energy and its existence prevents gluon densities from 
rapid growth. Existing data suggest that the phenomenon of saturation occurs in nature. The seminal example is provided by a 
discovery of the geometrical scaling in HERA data \cite{Stasto:2000er} and more recently by successful description of the observed de-correlation of forward di-hadrons in d+Au collisions as compared to p-p collisions in the RHIC data \cite{Braidot:2010ig,Braidot:2011zj,Adare:2011sc,Meredith:2009fp}. 
\begin{figure}[t!] 
\includegraphics[width=7.5cm]{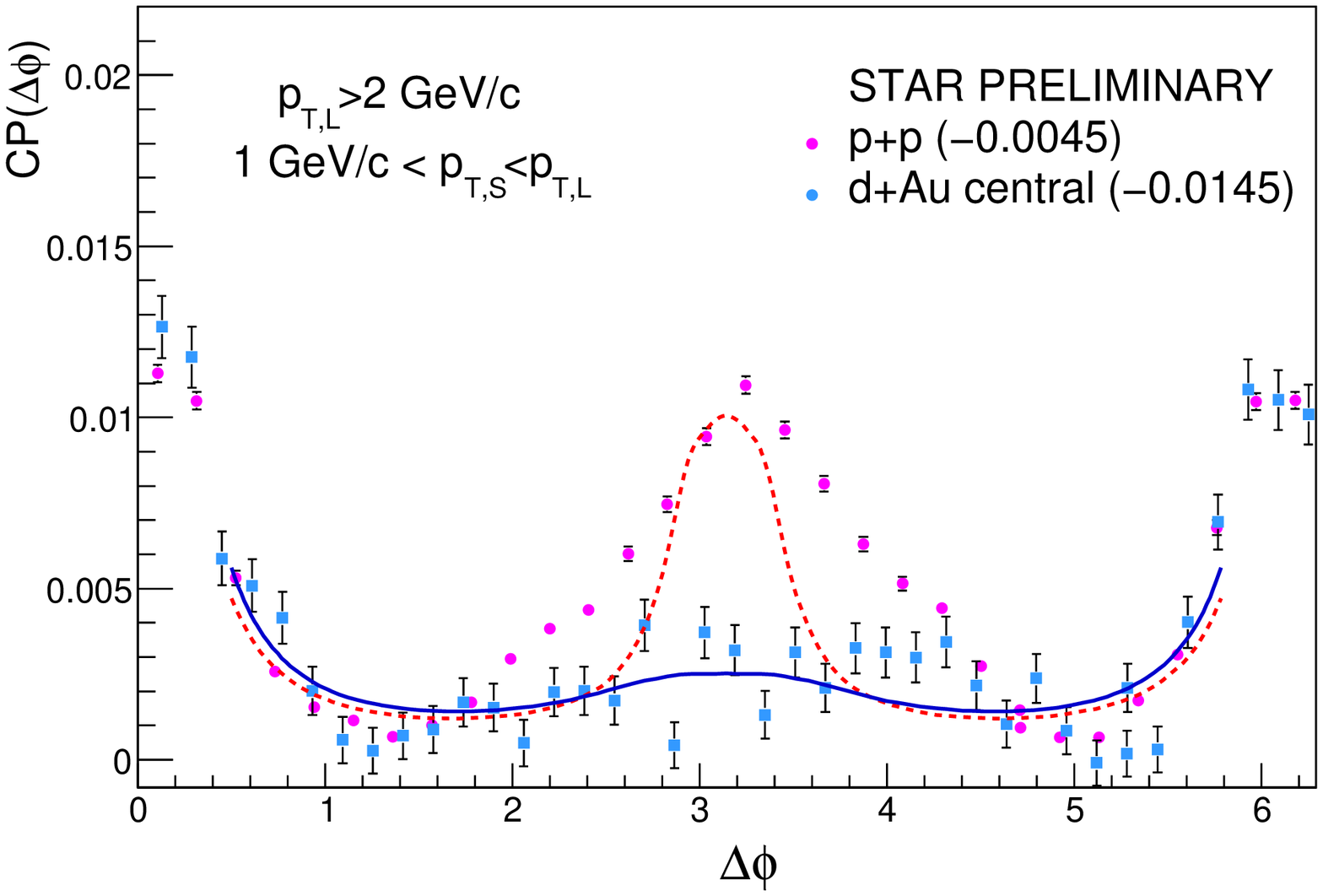}
\hfill
\includegraphics[width=7.5cm]{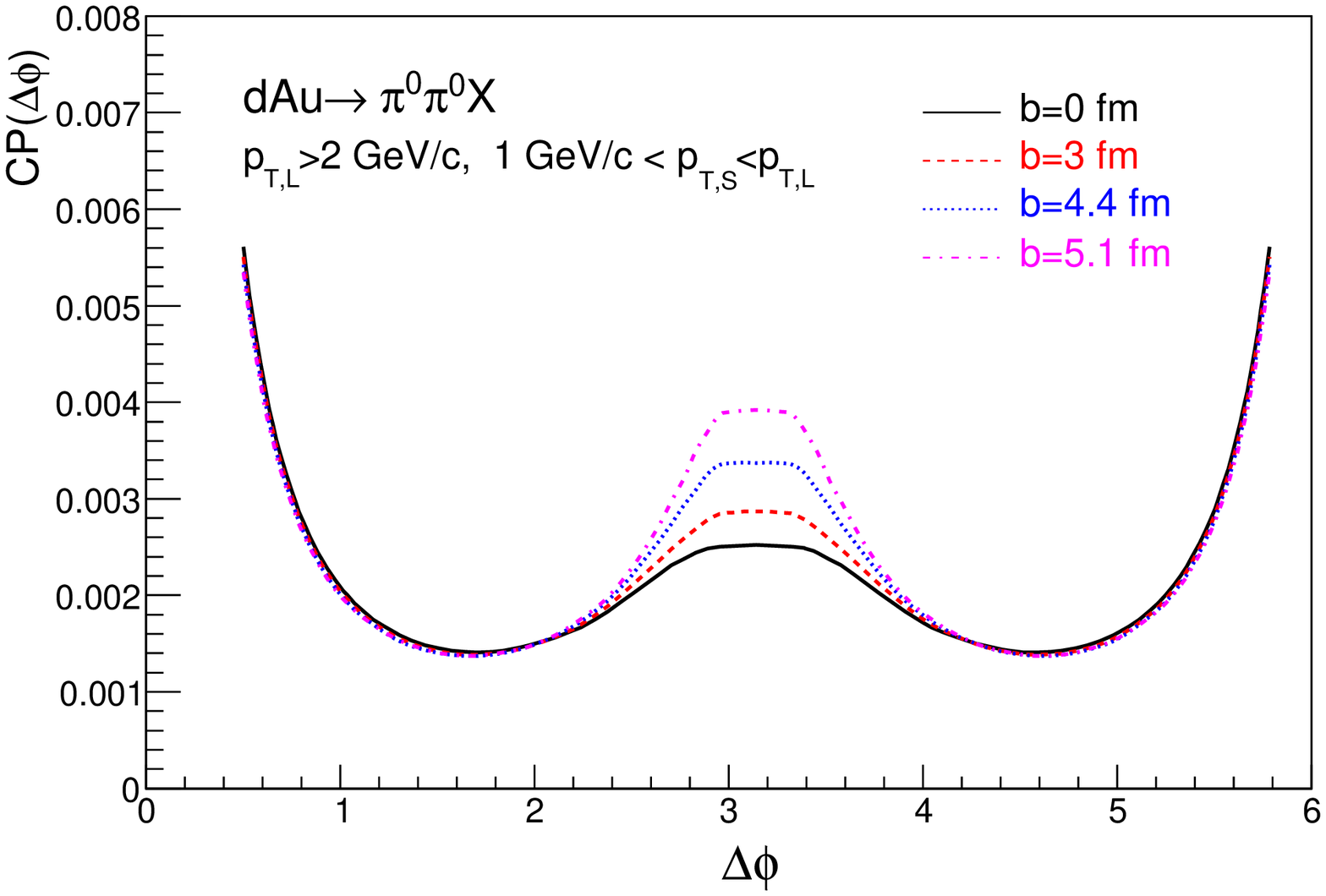}
\caption{The coincidence probability at a function of $\Delta\phi$. Left: CGC calculations \cite{arXiv:1005.4065} for p+p and 
central d+Au collisions, the disappearance of the away-side peak is quantitatively consistent with the STAR data. Right: CGC
 predictions for different centralities of the d+Au collisions, the near-side peak is independent of the centrality, while the 
 away-side peak reappears as collisions are more and more peripheral.}
\label{fig:saturation}
\end{figure}

Here we will focus  on the second of process mentioned above.
The kinematic range for forward particle detection at RHIC is $x_p\!\sim\!0.4$ and 
$x_A\!\sim\!10^{-3}$ at the $\sqrt{s}=200$ GeV. Therefore the dominant partonic subprocess is initiated by valence quarks in deuteron, and the $dAu\!\to\!h_1h_2X$ cross-section is obtained from the $qA\to qgX$ cross-section.
   \cite{arXiv:1005.4065}.
The de-corelation is measured as the coincidence probability to, given a trigger particle in a certain momentum range, produce
 an associated particle in another momentum range. It is given by:
\begin{equation}
CP(\Delta\phi)=\frac{N_{pair}(\Delta\phi)}{N_{trig}}\
\end{equation}
with
\begin{equation}
N_{pair}(\Delta\phi)=\int\limits_{y_i,|p_{i\perp}|}
\frac{dN^{pA\to h_1 h_2 X}}{d^3p_1 d^3p_2}\ ,\quad
N_{trig}=\int\limits_{y,\ p_\perp}\frac{dN^{pA\to hX}}{d^3p} .
\end{equation}
In Fig.~\ref{fig:saturation} the comparison of data to CGC predictions is shown. The agreement with data is very good and provides support for 
alternative explanations of the phenomenon of suppression of the forward hadron in d+Au which might also be explained by partonic energy 
loss during propagation through nuclear matter and which is not considered in the CGC framework. 
\subsection{Towards p-Pb in LHC}
\begin{figure}[t!]
  \begin{picture}(30,50)
    \put(-150, -210){
      \includegraphics{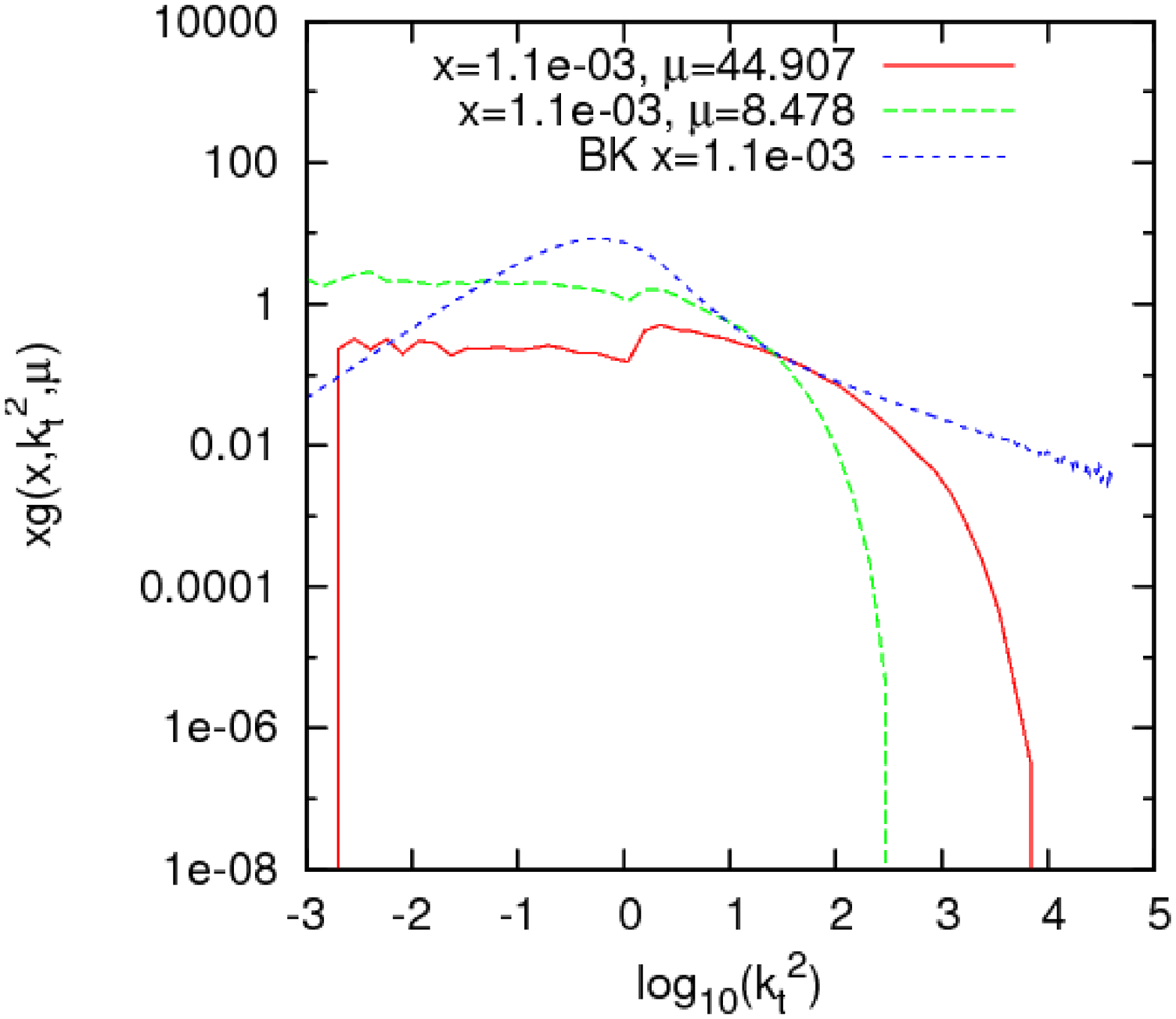}
    }

     \end{picture}
\vspace{5cm}
\caption{\em \small Comparison of unintegrated gluon densities obtained from rc-BK from \cite{arXiv:1012.4408} and CCFM evolution \cite{arXiv:1008.0152}. The blue dotted line corresponds to the rc-BK gluon density while the red and green lines correspond to the CCFM unintegrated gluon densities evaluate at different factorization scales.}
\label{fig:glueplots}
\end{figure}

One of the prospects for the LHC is to study collisions of protons with Pb nuclei. For a detailed
 prospect for this we refer the reader to \cite{arXiv:1012.4408}. One of the opportunities is to reduce the ucnertainity of nuclear pdfs and use
  this knowledge to interpret $A+A$ data. From our perspective the most interesting opportunity is to make use of the asymmetry of this initial
   configuration to access the forward region and to study the saturation effects with nuclear targets extending the kinematically 
   access by several orders of magnitude in $x$. To be prepared for this experiment and fully profit from it, one needs however
    further theoretical developments in the field of unintegrated parton densities. One of the open questions is how to account 
    both for exclusive final states at large $p_T$ like for example discussed above di-jets and saturation. The problem is that
     the frameworks used at present either accounts for hardness like DGLAP, CCFM or for saturation like CGC/BK and an approach when one accounts for both is missing. Practically it means that the CCFM gluon density is not proper at small $k_T$ while the BK gives unphysical gluon densities at large $k_T$, see Fig.~\ref{fig:glueplots}). It is however possible to introduce nonlinear effects into the CCFM framework and therefore to allow both for hard processes and nonlinear effects within one framework. One of the methods is to introduce proper boundary conditions which suppress the gluon density for certain combinations of $k_T$ and $x$ \cite{arXiv:0812.4082,arXiv:0906.2683,arXiv:1005.5153}. Another approach is to allow for coherent emission of gluons accompanied by the dynamical fusion of gluons \cite{arXiv:1111.6928}.  
\label{sec:trv}
\section{Conclusions}
We gave an overview of forward physics at LHC and RHIC focusing on jet final states. The goal was to show frameworks which 
follow from the same perturbative  principle at high energy asymptotics i.e.\ the BFKL like summation of large logarithms of the type 
$\alpha_s^m\ln(s/s_0)^n$. We also pointed at the possible prospects for p+Pb 
collisions which would provide an excellent opportunity to constrain parton densities and to study both hard processes and dense 
systems. This requires development of a unified framework which will allow 
for studies of parton saturation and hard processes.      
\section*{Acknowledgments}
I would like to thank the organizers and in particular Kazunori Itakura and Javier Albacete for the invitation and the excellent atmosphere of the meeting. I would also like to thank for partial financial support provided by organizers. Also the financial support of Fundacja na rzecz Nauki Polskiej with grant Homing Plus/2010-2/6 is acknowledged. 
Finally I acknowledge the discussions with collaborators M. Deak, F. Hautmann, H. Jung and S. Sapeta and comments on the manuscript by A. van Hameren.

\end{document}